# ELECTROMECHANICAL RELIABILITY TESTING OF THREE-AXIAL SILICON FORCE SENSORS

S. Spinner[1,2], J. Bartholomeyczik[1], B. Becker[2], M. Doelle[1],
O. Paul[1], I. Polian[2], R. Roth[3], K. Seitz[3], and P. Ruther[1]

[1]Department of Microsystems Engineering (IMTEK), University of Freiburg, Germany
[2]Institute for Computer Science, University of Freiburg, Germany
[3]Carl Zeiss, Industrielle Messtechnik GmbH, Oberkochen, Germany

## ABSTRACT

This paper reports on the systematic electromechanical characterization of a new three-axial force sensor used in dimensional metrology of micro components. The silicon based sensor system consists of piezoresistive mechanical stress transducers integrated in thin membrane hinges supporting a suspended flexible cross structure. The mechanical behavior of the fragile micromechanical structure is analyzed for both static and dynamic load cases. This work demonstrates that the silicon microstructure withstands static forces of 1.16 N applied orthogonally to the front-side of the structure. A statistical Weibull analysis of the measured data shows that these values are significantly reduced if the normal force is applied to the back of the sensor. Improvements of the sensor system design for future development cycles are derived from the measurement results.

## 1. INTRODUCTION

The reliability of microelectromechanical systems (MEMS) receives growing attention since MEMS are employed in safety-critical fields such as automotive, aerospace and medical applications. The reliability testing of MEMS devices includes (i) ensuring the correct functionality of the system after fabrication, i.e., the manufacturing test and (ii) investigation of possible deterioration of the device during application. The second issue is of special importance as MEMS applications often intrinsically involve mechanical stress which could damage or destroy the microstructure or lead to an unexpected behavior of the system [1].

The focus of manufacturing tests is to decide whether a system or component is 'good' and can be shipped to the costumer or not. In high-volume manufacturing the distinction between 'good' and 'bad' devices is done using electrical test methods. This is due to the fact that these methods are faster than optical or chemical techniques, and consequently are less expensive. Moreover, available automatic test equipment for integrated circuits (ICs) can be leveraged [2].

The reliability of MEMS is not only an important issue in safety-critical applications but also in measurement applications. For example, the ongoing miniaturization of mechanical and optical components requires precise and flexible tools for three-dimensional coordinate measurements [3-5]. Core components of these measurement systems are highly sensitive tactile sensors capable of minimizing contact forces. The correct operation of the total measurement system is strongly connected with the reliability of the tactile sensor itself. If the sensor degrades, the accuracy of the measurement results decreases.

Section 2 contains a description of the 3D force sensor which is the device under test (DUT) of this reliability study. The sensor behavior under applied normal forces and results of a finite element simulation of the sensor are reported. A short overview on the used measurement system is given in Section 3. This measurement system is able to induce stress to a DUT with simultaneous electrical measurement. Section 4 describes the performed measurements and respective results, e.g., fracture load and stiffness of the force sensor. A short description of how the sensor could be improved in terms of reliability is given in Section 5.

## 2. FORCE SENSOR

The three-axial force sensor shown in Figure 1 consists of a flexible cross structure realized using deep reactive ion etching of single crystal silicon [3]. The arms of the cross structure are connected to a silicon frame and to the centeral square of the cross through thin silicon membrane hinges with a thickness of 25 μm. The overall in-plane dimensions of the cross are $4.5 \times 4.5$ mm$^2$. A probe pin made of stainless steel with a length 7 mm and carrying a probe sphere is mounted to the center of the silicon cross. It serves as the tactile element of the force sensor. Forces applied to the tac-tile element deform the cross structure as schematically shown in Figure 1 (d) for forces along and orthogonal to the probe pin. This deformation is detected using piezoresistors implanted in the thin membrane hinges. Each cross arm contains four p-doped resistors connected in a Wheatstone bridge configuration [3]. The piezoresistors with a length to width ratio $L/W = 2$ are oriented parallel or orthogonal to the <110>-directions of the silicon crystal. Due to the selected p-type doping and the respective orientation of the piezoresistors, the stress sensitivity to normal stress components $\sigma_{xx}$ and $\sigma_{yy}$ is optimized.

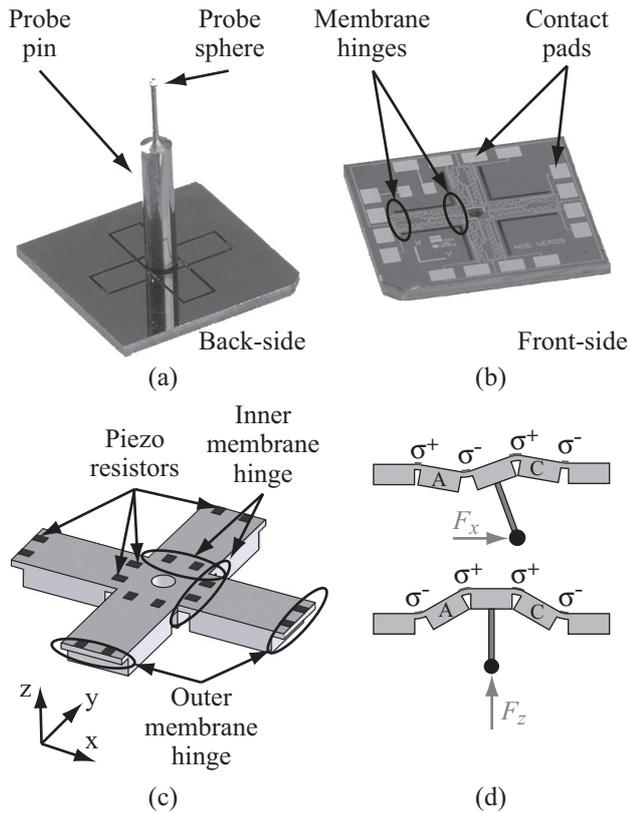

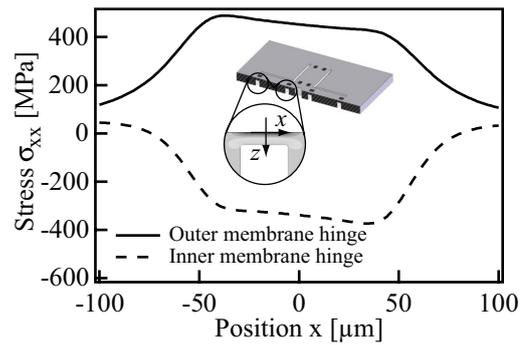

*Figure 1: Three-axial force sensor; (a) sensor back-side with assembled probe pin, (b) sensor front-side with contact pads, (c) schematic view of the flexible cross structure from the front-side indicating the position of the piezoresistors, (d) principle of operation applying forces in the x and z directions.*

The relative change $\Delta\rho/\rho_0$ of the resistivity $\rho$ of the piezoresistors is given by [6]

$$\frac{\Delta\rho}{\rho_0} = \pi_l \sigma_l + \pi_t \sigma_t, \quad (1)$$

where $\rho_0$, $\pi_l$, $\pi_t$, $\sigma_l$ and $\sigma_t$ denote the resistivity of the doped material in the stress free state, the longitudinal and transversal piezoresistive coefficients and the respective stress components parallel and transverse to the orientation of the piezoresistors, respectively. For the given doping of the resistors, the piezoresistive coefficients are on the order of $\pi_l = 71.8\times10^{-11}$ Pa$^{-1}$ and $\pi_t = -66.3\times10^{-11}$ Pa$^{-1}$ [7].

*Figure 2: Distripution of the $\sigma_{xx}$ stress component on the surface of the membrane hinges. The stress values are extracted from a 2D simulation. The position $x = 0$ indicates the respective centers of both membranes.*

Assuming homogeneous stress distributions within each membrane hinge, the offset voltages $V_{off}$ of the Wheatstone bridges are given by

$$V_{off} = \frac{\Delta\rho_{in} - \Delta\rho_{out}}{2\rho_0 + \Delta\rho_{in} + \Delta\rho_{out}} V_{ges} \quad (2)$$

where $V_{ges}$, $\Delta\rho_{in}$ and $\Delta\rho_{out}$ denote the supply voltage of the bridge, and resitivity changes on the inner and outer membrane hinges, respectively.

### 2.1. FEM Simulations

The mechanical behavior of the three-axial force sensor has been evaluated with 2D and 3D finite element (FEM) simulations using ANSYS 9.0. The simulations are used to extract the stress distribution in the thin membrane hinges under various load cases as well as typical force-displacement curves. As an example, Figure 2 shows the $\sigma_{xx}$ stress component on the chip surface of the left outer and left inner membrane hinge applying a normal force of $F_z = 0.5$ N to the front-side of the chip. While the inner membrane hinge is compressed with $-373$ MPa, the top surface of the outer membrane is under tensile stress with 489 MPa. Applying the same normal load in the opposite direction, this stress distribution is just reversed.

### 3. EXPERIMENTAL SETUP

The experimental setup shown in Figure 3 consists of three major components: (i) a positioning platform, (ii) the mechanical impact control unit and (iii) a mounting frame to which the impact control unit is attached. The positioning platform comprises two x- and y-translation stages, a z-stage, a rotation stage and a 6-inch vacuum chuck. With a maximum transverse path length of 200 mm of the x- and y-translation stages the setup enables device testing on the wafer level using 6-inch wafers. The positioning accuracy of the $x$- und $y$-stages is specified to 2 μm. The rotation stage enables the alignment of the device under test (DUT) with respect to the mechanical impact control unit.

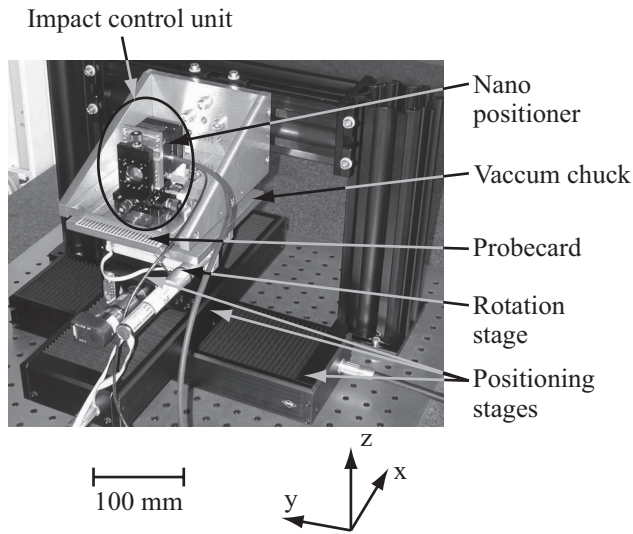

*Figure 3: Experimental setup for the electromechanical characterization of MEMS components, e.g. a three-axial force sensor.*

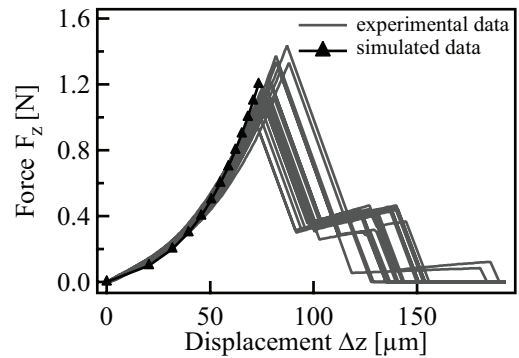

*Figure 4: Force versus displacement curves applying normal forces $F_z$ to the sensor front; discontinuities indicate failure of membrane hinges (straight lines: experimental data from 20 sensors; symbols: simulated data).*

The mechanical impact control unit consists of (i) an $xyz$-nanopositioning system (NanoCube from PI) with traverse path lengths of 100 μm and positioning accuracies of 20 nm for each direction, (ii) a force sensor with a resolution of 5 mN and (iii) a probecard for electrical contacting of the DUT. The nanopositioning system in combination with the force sensor enables the laterally controlled application of normal forces $F_z$ up to 3.6 N at frequencies as high as 20 Hz. Detailed information on the experimental setup is given in [8].

## 4. RESULTS

### 4.1. Static measurement

To determine the stiffness of the three-axial force sensor under normal loads $F_z$, sensor dies were driven against the vertically fixed force sensor using the $z$-stage of the positioning platform. The maximum displacement $\Delta z_{max}$ of the $z$-stage was limited to 200 μm sufficient to break at least one membrane hinge per cross arm. Figure 4 shows as an example displacement-force curves of 20 different force sensors applying the forces to the sensor front. For comparison, a simulated displacement-force curve for a nominal membrane thickness of 25 μm is added to Figure 4 indicating the excellent match of FE simulation and experiment. Discontinuities of the curves indicate the failure of individual membrane hinges. Similar results are obtained applying normal forces from the back-side of the sensor.

From the experimental data in Figure 5 three specific regions of the curves can be distinguished. For displacements below 20 μm, the force sensors show a linear response to normal forces $F_z$ applied to either the front or back-side. For larger displacements $\Delta z$ a non-linear behavior of force versus displacement can be observed. Finally, the applied stress exceeds the fracture toughness of the membrane hinges and the failure of individual hinges can be observed.

Extracting the slopes of the displacement-force curves in Figure 4 within region 1 at a maximum displacement $\Delta z = 20$ μm, an initial stiffness of $(7.01 \pm 0.57)$ mN/μm is obtained applying normal forces to the sensor front. In case of normal loads applied to the back-side, a slope of $(6.61 \pm 0.23)$ mN/μm is extracted.

The sensors were found to withstand on average normal loads applied to the front-side of $F_{z,ave} = (1.16 \pm 0.12)$ N before one of the membrane hinges fails. The corresponding displacements of the flexible cross structure are $\Delta z_{ave} = (78.2 \pm 4.6)$ μm. In case of normal loads applied to the back-side, these values reduce to $F_{z,ave} = (0.72 \pm 0.11)$ N and $\Delta z_{ave} = (55.1 \pm 4.9)$ μm, respectively.

The statistical analysis of the fracture loads is shown in Figure 5 for normal forces applied to the sensor front- and back-side. The measured minimal fracture loads are fitted to a Weibull density function $p(F)$ [5]

$$p(F) = 1 - \exp\left(-\frac{F}{F_0}\right)^\beta \qquad (3)$$

where $F_0$ and $β$ denote fit parameters. The Weibull curves indicate that the fracture load for forces applied to the front-side is significantly higher than that for normal loads to the back-side.

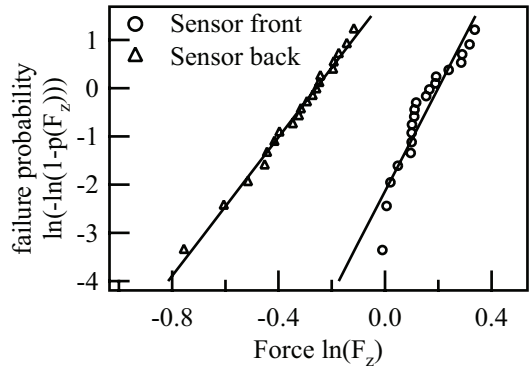

*Figure 5: Statistical analysis of the fracture for forces applied load in the z-direction; triangles and circles represent the measurement results applying a force to the back and front of the sensor, respectively.*

The fit parameters $F_0$ and $\beta$ are summarized in Table 1 together with the *R*-parameter calculated using

$$R = \frac{\sum y_i^2 - \sum (y_i - y_i')^2}{\sum y_i^2} \quad (4)$$

where $y_i$ and $y_i'$ denote the estimated failure probability from the measurement results and the failure probability calculated from the Weibull density function using the extracted fit parameters $F_0$ and $\beta$. The *R*-parameter indicates the quality of the Weibull fit.

|  | Front-side | Back-side |
|---|---|---|
| $F_0$ [N] | 1.22 | 0.77 |
| $\beta$ [-] | 10.69 | 7.21 |
| $R$ [-] | 0.9756 | 0.9983 |

*Table 1: Fit parameters $F_0$ and $\beta$ for the Weibull density function and the R-parameter given in Eq. (4).*

From the fitted Weibull density function it is possible to calculate the expected failure probability of sensors loaded with a normal force $F_z$ or driven to displacement $\Delta z$. It turned out from the statistical analyses that a failure probability of 10 ppm is achieved for normal forces $F_{z, max}$ = 0.42 N applied to the front-side. In case of normal forces applied to the back-side, this maximum force $F_{z, max}$ is decreased to 0.16 N. Table 2 summarizes values of the maximum normal force $F_{z, max}$ and maximum displacement $\Delta z_{max}$ for some selected failure rates.

Optical inspections of the force sensors during the destructive mechanical testing show that the membrane hinges typically fail as expected from the FE simulations. Applying normal forces $F_z$ to the front-side of the sensors, in most cases the outer membrane hinges failed first. In con-trast, when normal forces are applied to the back-side, typically the inner membrane hinges failed first. After completion of the static measurements, the number of broken membrane hinges per sensor are counted and summarized in Table 3 for 20 tested sensor chips per direction in which the normal fores are applied. It is obvious, that by ap-plying normal forces to the sensor front mainly the outer membrane hinges fail. Thus, in most cases the flexible cross structure is broken out of the silicon frame. In contrast to this, mainly the inner hinges fail when the normal force is applied from the back-side. In this case, the central part of the cross is detached.

| Failure probability [ppm] | Front-side | | Back-side | |
|---|---|---|---|---|
|  | $F_{z, max}$ [N] | $\Delta z_{max}$ [μm] | $F_{z, max}$ [N] | $\Delta z_{max}$ [μm] |
| 1 | 0.34 | 39.38 | 0.11 | 19.35 |
| 10 | 0.42 | 44.35 | 0.16 | 23.19 |
| 100 | 0.52 | 49.95 | 0.21 | 27.80 |

*Table 2: Calculated maximum normal force $F_{z, max}$ and maximum displacement $\Delta z_{max}$ using the Weibull density function in Eq. (3) for a given failure probability.*

| Normal force $F_z$ applied to | Membrane hinges | |
|---|---|---|
|  | inner | outer |
| Front-side | 19 | 76 |
| Back-side | 68 | 25 |

*Table 3: Counting result of the failed membrane hinges after the load cycle of the 20 sensors for each side.*

Figure 6 shows typical offset signals $V_{off, A}$ to $V_{off, D}$ of the four Wheatstone bridges A to D and the applied normal force $F_z$ as a function of the displacement $\Delta z$. The discontinuities in the force $F_z$ and offset signal $V_{off}$ versus displacement $\Delta z$ indicate the failure of one of the membrane hinges. From the electrical offset signals the order in which cross arm failure occurs can be extracted. In the example shown in Figure 6, first cross arm B fails (see (1) in Figure 6) followed by cross arm C (see (2)). As the cross arm C carries the supply leads of the stress sensor, offset measurements are not possible after failure of this arm. However, measuring the electrically accessible resistors only, we conclude that in this example the membrane failure occured in the outer hinges as expected from the direction of load application.

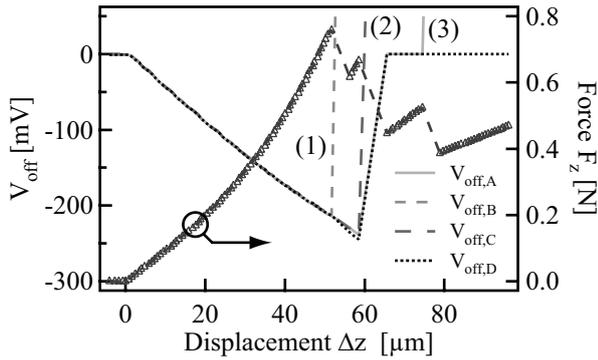

*Figure 6: Sensor bridge signals $V_{off,A}$ to $V_{off,D}$ and normal force $F_z$ applied to the front-side versus displacement $\Delta z$ of the cross structure center; discontinuities in $V_{off}$ and $F_z$ indicate the failure of membrane hinges (failure of a membrane hinge (1) B, (2) C, and (3) A).*

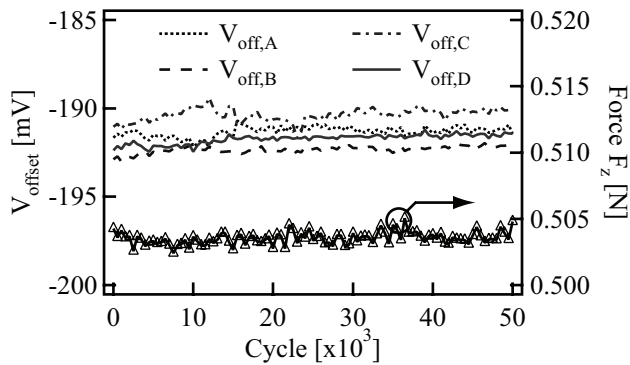

*Figure 7: Sensor bridge signals $V_{off,A}$ to $V_{off,D}$ and normal force $F_z$ applied to the front-side of the device during 50 000 load cycles; $V_{off}$ measured at a applied normal force $F_z$ of 0.5 N and a supply voltage $V_{ges}$ of 1 V.*

### 4.2. Dynamic Measurement

In the considered long term testing sequence the applied normal force $F_z$ was varied between 0.01 N and 0.5 N at a frequency of 2 Hz. This causes a maximum deformation $\Delta z$ of the cross structure of ca. 40 µm. As an example, Figure 7 shows the bridge offset voltages $V_{off,A}$ through $V_{off,D}$ of the four Wheatstone bridges A to D at a fixed applied normal force $F_z$ = 0.5 N during 50 000 load cycles. The offset voltages $V_{off}$ were measured after intervals of 500 load cycles at a supply voltage of $V_{ges}$ = 1 V. Prior to the offset measurements, the experimental setup was reset for each measurement to ensure high quality results. The average offset voltage of the four Wheatstone bridges and the applied load of the long term measurement are summarized with the respective standard deviations in Table 4. With a standard deviation of the applied load better 0.1% and below 0.2% of the measured offset voltages, it can be concluded that no significant degradation of the force sensor is observable within the 50 000 load cycles.

| | Average | Standard deviation |
|---|---|---|
| Force $F_z$ [N] | 0.50352 | 0.00037 |
| $V_{off,A}$ [mV] | -191.32 | 0.29 |
| $V_{off,B}$ [mV] | -192.33 | 0.19 |
| $V_{off,C}$ [mV] | -190.36 | 0.36 |
| $V_{off,D}$ [mV] | -191.73 | 0.26 |

*Table 4: Measurement results over the 50 000 applied load cycles.*

### 5. CONCLUSIONS

Static and dynamic reliabilty testing of a silicon based three-axial force sensor used in dimensional metrology has been conducted. From the static measurements it can be concluded that the sensor reliability can significantly be improved mounting the probe pin to the front- instead of backside of the sensor die. By this design change, the average fracture load is increased by a factor of 1.6 from 0.72 N to 1.16 N. As indicated by the statistical analyses, to limit the failure probability of 1 ppm a maximum normal force $F_{z,max}$ = 0.34 N applied to the front-side is tolerable. In contrast, with the the probe pin mounted to the back-side, the tolerable force $F_{z,max}$ is reduced to 0.11 N.

In case of dynamic long term measurements with a maximum normal force of 0.5 N it could be shown that the sensor experiences no observable degradation within 50 000 load cycles. With standard deviations of the measured offset voltage below 0.2% these changes are within the measurement accuracy of the experimental setup.

The experiments revealed that the tree-axial force sensor is a highly reliable microsystem. As the typical sensor displacements in metrological applications are below 2 µm during surface detection, the observed maximum displacements of 38.3 µm are equivalent to a 19-fold overload in terms of displacement. This corresponds to a 22-fold overload in terms of force levels.

### 6. ACKNOWLEDGEMENT

This work was supported by the DFG grant GRK 1103/1 of the Deutsche Forschungsgemeinschaft.